\documentclass[12pt, preprint]{revtex4-1}
\usepackage{graphicx}
\usepackage[OT1]{fontenc}
\usepackage[utf8]{inputenc}
\usepackage{bm,amsmath,amssymb,amsbsy,amstext}
\usepackage[colorlinks=false]{hyperref}
\usepackage{xcolor}

\begin{document}

\title{A generic method for equipping arbitrary rf discharge simulation frameworks with external lumped element circuits}
\author{Frederik Schmidt}
\affiliation{%
Institute of Theoretical Electrical Engineering, Ruhr University Bochum, Bochum, 44780, Germany
}%

\author{Jan Trieschmann}
\affiliation{%
Electrodynamics and Physical Electronics Group, Brandenburg University of Technology Cottbus-Senftenberg, Cottbus, 03046, Germany
}%

\author{Tobias Gergs}
\affiliation{%
Electrodynamics and Physical Electronics Group, Brandenburg University of Technology Cottbus-Senftenberg, Cottbus, 03046, Germany
}%

\author{Thomas Mussenbrock}
\affiliation{%
Electrodynamics and Physical Electronics Group, Brandenburg University of Technology Cottbus-Senftenberg, Cottbus, 03046, Germany
}%
\date{\today}

\begin{abstract}

\medskip

External electric circuits attached to radio-frequency plasma discharges are essential for the power transfer into the discharge and are, therefore, a key element for plasma operation. Many plasma simulations, however, simplify or even neglect the external network. This is because a solution of the circuit's auxiliary differential equations following Kirchhoff's laws is required, which can become a tedious task especially for large circuits. This work proposes a method, which allows to include electric circuits in any desired radio-frequency plasma simulation. Conceptually, arbitrarily complex external networks may be incorporated in the form of a simple netlist. The suggested approach is based on the harmonic balance concept, which splits the whole system into the nonlinear plasma and the linear circuit contribution. A mathematical formulation of the influence of the applied voltage on the current for each specific harmonic is required and proposed. It is demonstrated that this method is applicable for both simple global plasma models as well as more complex spatially resolved Particle-in-Cell simulations.

\end{abstract}

\maketitle

\section{\label{sec:introduction} Introduction}

Radio frequency plasma sources such as capacitively or inductively coupled plasmas (CCPs/ICPs) are necessarily operated using external electric circuits to transfer the generated power into the discharge \cite{lieberman_principles_2005, chabert_physics_2011}. These circuits include generators, matching networks, filters and power lines among others. The voltage waveform at the driven electrodes and the current flowing through the plasma discharge depend decisively on the electrical properties of the network elements, which the plasma interacts with. Plasmas operated at radio frequency and low pressure also show a nonlinear behavior, making the interaction not easily predictable \cite{mussenbrock_nonlinear_2007,mussenbrock_enhancement_2008, czarnetzki_self-excitation_2006, lieberman_effects_2008, miller_electrical_1992, ziegler_temporal_2009, yamazawa_effect_2009, yamazawa_electrode_2015}.

Many plasma simulation techniques focus on the plasma dynamics itself and external circuits are often neglected or drastically simplified, e.g., to a simple bias capacitance. Electrical equivalent circuit models allow to comprisingly incorporate the plasma and complex external circuits in the solution algorithms \cite{schmidt_consistent_2018, schmidt_multi_2018}. In contrast, Particle-in-Cell (PIC) simulations have been coupled to external series circuits consisting of a resistance, an inductance and a capacitance via conservation of charge by Verboncoeur et al.\ \cite{verboncoeur_simultaneous_1993}. In their method, the differential equations following Kirchhoff's circuit laws have been discretized and solved, ensuring numerical stability of the whole simulation. While conceptually possible, an extension to more complex external networks involves a similarly elaborate procedure, which limits the practical applicability.

In this work, we propose a method for coupling an external circuit to any desired plasma simulation that gives a voltage-current relation based on the method of harmonic balance \cite{maas_nonlinear_2003, gilmore_nonlinear_1991}. Once implemented, the external circuit can be included via a simple netlist, hence, making changes in the setup easy to implement and investigate. The main idea is to split the whole system into a linear part -- the electric circuit -- and a nonlinear part -- the plasma. Notably, the nonlinear part may contain linear elements, whereas the linear part must not contain any nonlinear elements. The voltage drop between the interconnection(s) connecting the two is sought for, so that the currents nullify one another, i.e. Kirchhoff's nodal law is satisfied. This needs to be accomplished for every harmonic of interest, hence the term harmonic balance. The procedure is detailed subsequently.

The manuscript is organized as follows: In chapter \ref{sec:method} the principle and the algorithm of the method is discussed. Two different plasma descriptions are coupled to external circuits in this work. On the one hand, a global equivalent circuit is utilized, which can also be simulated using SPICE for reference. On the other hand, a 1-dimensional Particle-in-Cell (PIC) simulation is used and coupled to two variants of external circuits. These methods are proposed in detail in chapter \ref{sec:model} and the respective results discussed in comparison with reference methods in chapter \ref{sec:results}.

\section{\label{sec:method} Harmonic Balance Algorithm}

Harmonic balance is a common method for calculating the interactions of linear circuits with nonlinear elements such as diodes or transistors \cite{maas_nonlinear_2003, gilmore_nonlinear_1991}. The fundamental idea is to split the circuit into a linear and a nonlinear part and find the voltage in between those regimes for which each harmonic in the respective currents is the same as in the other circuit, i.e., balanced. Hence the term harmonic balance.

A detailed description of the implementation is provided by Maas \cite{maas_nonlinear_2003}. In this chapter, the basic concept is reviewed and details about the necessary changes in the method to adapt it for plasma--circuit simulations are discussed.

\begin{figure}[t!]
	\centering
	\resizebox{8cm}{!}{		
		\includegraphics[width=8cm]{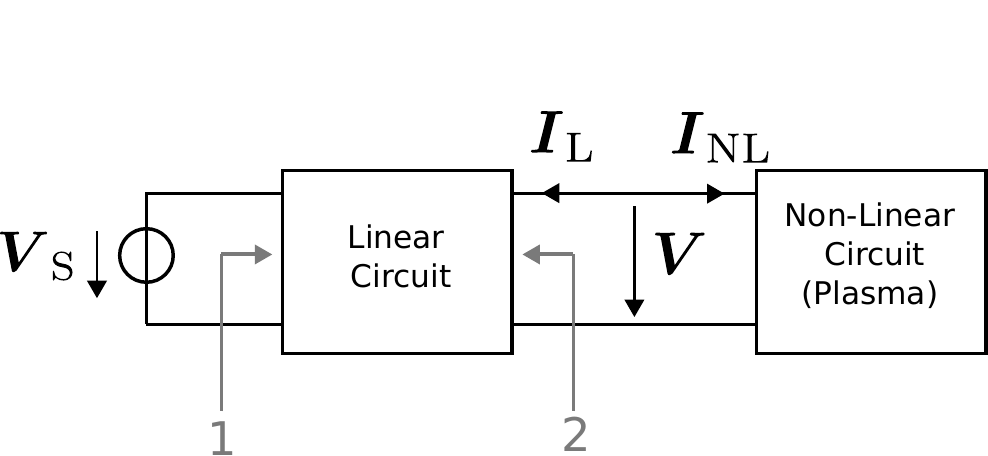}
		
	}
	\caption{Harmonic balance setup}
	\label{fig:harmonic_balance}
\end{figure}

The linear circuit consisting of resistances, inductances and capacitances on the one hand and the nonlinear plasma on the other hand. This is depicted in figure~\ref{fig:harmonic_balance}. The transient voltage $v(t)$ dropping between the interconnection and, respectively, the transient current $i(t)$ entering the nonlinear circuit part can be written in Fourier series representation as
\begin{eqnarray}
\label{eq:fourier_series1}
i(t) &=&  \sum\nolimits_{k=-K}^K I_k e^{j k \omega t},\\
\label{eq:fourier_series2}
v(t) &=&  \sum\nolimits_{l=-K}^K V_l e^{j l \omega t},
\end{eqnarray}

with $I_k = I_{-k}^*$ and $V_l = V_{-l}^*$. For the purpose of practicability $K$ is chosen as a finite number. In the algorithm described in the following paragraphs, only values for $k,l \geq 0$ are considered, which reduces the complexity of the procedure. Thereby (temporarily) complex values for $i(t)$ and $v(t)$ arise.

The Fourier coefficients $I_k$ and $V_l$ can be consistently defined as
\begin{eqnarray}
\label{eq:coefficients_complex}
I_k &=& \frac{1}{T} \int_{0}^T i(t) e^{- j k \omega t} dt, \\
V_l &=& \frac{1}{T} \int_{0}^T v(t) e^{- j k \omega t} dt.
\end{eqnarray}
These coefficients can be written in vector form $\boldsymbol{I}_\mathrm{NL}$ and $\boldsymbol{V}$ of dimension $K+1$, containing a (real valued) DC entry, a complex valued fundamental frequency component and $K-1$ complex valued harmonics of the fundamental frequency. The goal in harmonic balance is to find a voltage $\boldsymbol{V}$ for which the current flowing into the linear circuit $\boldsymbol{I}_\mathrm{L}$ and the current flowing into the plasma $\boldsymbol{I}_\mathrm{NL}$ satisfy Kirchhoff's nodal law. In other words, a current error vector
\begin{align}
    \boldsymbol{F} = \boldsymbol{I}_\mathrm{L} + \boldsymbol{I}_\mathrm{NL}
	 \label{eq:ohms_law}
\end{align} 
can be defined which is desired to vanish, $\boldsymbol{F} = 0$. Note that also the linear current $\boldsymbol{I}_\mathrm{L}$ is written as a vector of dimension $K+1$.

The transadmittance matrix $\underline{\boldsymbol{Y}}$ of a linear circuit can be calculated using nodal analysis, which has to be done for each frequency. This can easily be automated and the linear circuit information thereby included via a simple netlist as it is done for example in SPICE~\cite{nagel_spice_1973}. This leads to a linear current entering port 2
\begin{align}
\label{eq:linear}
\boldsymbol{I}_\mathrm{L} = \underline{\boldsymbol{Y}_{21}} \cdot \boldsymbol{V}_\mathrm{S} + \underline{\boldsymbol{Y}_{22}} \cdot \boldsymbol{V},
\end{align}
which entails current contributions due to the voltages $\boldsymbol{V}_\mathrm{S}$ and  $\boldsymbol{V}$ at both ports 1 and 2 (cf.\ Figure~\ref{fig:harmonic_balance}). The linear current can be straightforwardly calculated in frequency space.

The plasma simulations used in this work are performed in time domain. Therefore, the nonlinear current $\boldsymbol{I}_\mathrm{NL}$ is calculated from the evolution $i(t)$, which is the result of a transient simulation of the plasma. The latter is subject to the voltage $v(t)$, obtained using the Fourier series representation of equation~\eqref{eq:fourier_series2} with coefficients $\boldsymbol{V}$.

\begin{figure}[t!]
	\centering
	\resizebox{8cm}{!}{		
		\includegraphics[width=8cm]{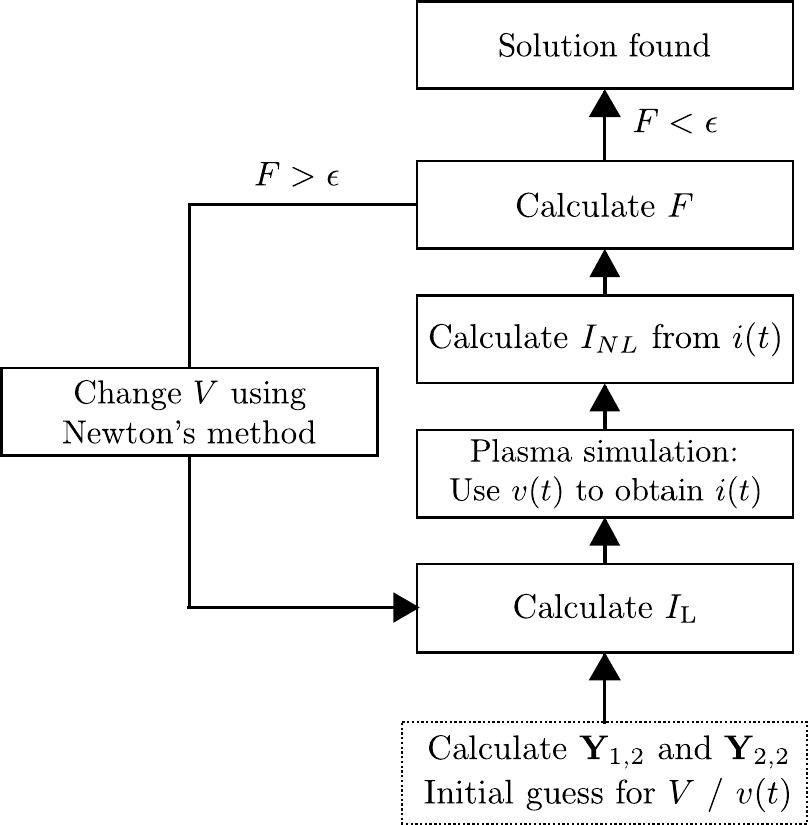}
		
	}
	\caption{Harmonic balance algorithm.}
	\label{fig:algorithm}
\end{figure}

The algorithm of the simulation is depicted in figure~\ref{fig:algorithm}. Starting with an initial guess of $\boldsymbol{V}$ all described values can be calculated. The most complex part of the algorithm is to change $\boldsymbol{V}$ until $\boldsymbol{F}$ is satisfactory small, which is done by utilizing Newton's method: After the $p$-th iteration step, the voltage $\boldsymbol{V}^{p+1}$ can be calculated using the previous value $\boldsymbol{V}^{p}$ and the error $\boldsymbol{F}$ in the form of
\begin{align}
\label{eq:newton_next_step}
	{\boldsymbol{V}}^{p+1} = \boldsymbol{V}^{p} - \underline{\boldsymbol{J}_\mathrm{F}}^{-1} {\boldsymbol{F}}({\boldsymbol{V}^p}),
\end{align}
with the Jacobian 
\begin{eqnarray}
	\underline{\boldsymbol{J}_\mathrm{F}} &=& \left.\frac{\partial {\boldsymbol{F}}({\boldsymbol{V}})}{\partial {\boldsymbol{V}}}\right|_{{\boldsymbol{V}} = {\boldsymbol{V}}^p} \nonumber \\
	\label{eq:jacobian}
	&=& \underline{\boldsymbol{Y}_{22}} + \frac{\partial \boldsymbol{I}_\mathrm{NL}}{\partial \boldsymbol{V}}.
\end{eqnarray}
The voltage source term $\underline{\boldsymbol{Y}_{21}} \cdot \boldsymbol{V}_\mathrm{S}$ is independent of $\boldsymbol{V}$ and vanishes.

The second term in equation~\eqref{eq:jacobian} can be written as
\begin{align}
	\frac{\partial I_k}{\partial V_l} = \frac{1}{T} \int_{0}^T \frac{  \partial i(t) }{\partial V_l} e^{- j k \omega t} dt.
\end{align}
This can be expanded to
\begin{align}
\label{eq:jacobian_element}
\frac{\partial I_k}{\partial V_l} =  \frac{1}{T} \int_0^T \frac{\partial i(t)}{\partial v(t)} e^{-j (k-l) \omega t} dt,
\end{align}
using $\partial v(t) / \partial V_l = e^{j l \omega t}$.~\cite{maas_nonlinear_2003}
For a time-varying voltage $v(t)$,  the entries of the Jacobian depend on
\begin{eqnarray}
\frac{\partial i(t)}{\partial v(t)} &=& \frac{\partial i(t)}{\partial t} \left( \frac{\partial v(t)}{\partial t} \right)^{-1} \nonumber \\
\label{eq:partial1}
&=& \frac{\sum\nolimits_{k=1}^K I_k k e^{j k \omega t}}{\sum\nolimits_{l=1}^K V_l l e^{j l \omega t} },
\end{eqnarray}
using the definitions from equations~\eqref{eq:fourier_series1} and \eqref{eq:fourier_series2}. For DC excitation, the time derivative of $v(t)$ vanishes and equation~\eqref{eq:partial1} cannot be utilized. In this case, it is safe to assume a linear relation of the current components to the voltage
\begin{eqnarray}
\frac{\partial i(t)}{\partial v^{(0)}}
\label{eq:partial_DC}
&=& \frac{\sum\nolimits_{k=0}^K I_k  e^{j k \omega t}}{V_0 }.
\end{eqnarray}
Equation \eqref{eq:partial1} and \eqref{eq:partial_DC} can be directly incorporated into equation~\eqref{eq:jacobian_element}.

By performing a plasma simulation with a specific voltage $\boldsymbol{V}$ (respectively $v(t)$, which may entail several non-zero components $V_l$), only the collective system response $\boldsymbol{I}_\mathrm{NL}$ to this particular excitation may be obtained. The system response is probed using $\frac{\partial i(t)}{\partial v(t)}$ for the specified work point only. Moreover, the integration kernel of equation~\eqref{eq:jacobian_element} solely depends on the index difference $k-l$ and not on the individual indices $k$ and $l$. Therefore, for a given response $\frac{\partial i(t)}{\partial v(t)}$, the matrix elements $G_{k,l} = \frac{\partial I_k}{\partial V_l}$ are the elements of a circulant matrix $\underline{\boldsymbol{G}}$ (fully specified by a vector with elements $g_k = G_{k,\,l=0}$). The corresponding Jacobian matrix, however, is insufficient as it does not entail the isolated influence of all specific frequency components $V_l$ of the voltage $\boldsymbol{V}$ on the system -- specifically, on the current $\boldsymbol{I}_\mathrm{NL}$.

To obtain the selective system response to the $m$-th harmonic of the voltage and to set up the corresponding parts of the Jacobian, the simulation needs to be performed not only with $\boldsymbol{V}$, but also with $K+1$ simulations using a voltage, which is disturbed at the $m$-th entry. We define a voltage $\boldsymbol{V_{\Delta}}$, which contains these disturbances for each frequency component. The latter should not be too large in order to not change the state of the system when performing the respectively different simulations, but large enough to be distinctive from noise. We found empirically that  $\boldsymbol{V_{\Delta}} = 1/100 \, \boldsymbol{V}$ is a good choice. Especially for the DC value the disturbance may need to be chosen larger, which heavily depends on the specific plasma model. 

The simulation is now performed $K+1$ times with a voltage $\boldsymbol{\hat{V}}$, which differs from $\boldsymbol{V}$ at the $m$-th entry by $V_{\Delta m}$, specifically $\hat{V}_l = V_l + V_{\Delta m} \delta_{l m}$ with $\delta_{lm}$ the Kronecker delta. The current resulting from the disturbed excitation is denoted by $\boldsymbol{\hat{I}}$. To now account for the variances in the current due to the specific difference in the excitation at the $m$-th entry in the voltage, equation~\eqref{eq:partial1} can be reformulated to

\begin{align}
\label{eq:partial2}
\frac{\partial i_{\Delta}^{(m)} (t)}{\partial v_{\Delta}^{(m)} (t)} = \frac{\sum\nolimits_{k=1}^K (I_k -\hat{I}_k ) k e^{j k \omega t}}{ \sum\nolimits_{l=1}^K (V_l -\hat{V}_l) l e^{j l \omega t} } = \frac{\sum\nolimits_{k=1}^K (I_k - \hat{I}_k) k e^{j k \omega t}}{- V_{\Delta m} m e^{j m \omega t} },
\end{align}

which includes the change of all harmonics in the current $k \in \left[1,K\right]$ due to the $m$-th probing harmonic in the voltage at frequency $m \omega$. The influence of the DC current component neglected in equation~\eqref{eq:partial2} is immaterial, as any contribution is integrated out in equation~\eqref{eq:jacobian_element}. Again, equation~\eqref{eq:partial2} is not defined for DC. In this case the evaluation needs to be based on equation~\eqref{eq:partial_DC} leading to
\begin{align}
\label{eq:partial_DC2}
\frac{\partial i_{\Delta}^{(0)} (t)}{\partial v_{\Delta}^{(0)} } = \frac{\sum\nolimits_{k=0}^K (I_k - \hat{I}_k) e^{j k \omega t}}{- V_{\Delta 0}}.
\end{align}

Plugging equation~\eqref{eq:partial2} and \eqref{eq:partial_DC2} into equation~\eqref{eq:jacobian_element}, the $m$-th column vector of the Jacobian matrix
\begin{align}
\label{eq:jacobian_element2}
G_{k,\,m} = \frac{\partial I_{\Delta k}}{\partial V_{\Delta m}} =  \frac{1}{T} \int_0^T \frac{\partial i_{\Delta}^{(m)} (t)}{\partial v_{\Delta}^{(m)} (t)} e^{-j (k-m) \omega t} dt
\end{align}
may be evaluated. The latter can be computed directly in case of DC as a function of the response in the respective harmonic 
\begin{align}
\label{eq:jacobian_element_dc}
G_{k,\,0} = \frac{I_k - \hat{I}_k}{- V_{\Delta 0}}.
\end{align}

 Finally, the Jacobian matrix
\begin{align}
\underline{\boldsymbol{J}_F} = \underline{\boldsymbol{Y}_{22}} +
	\begin{bmatrix}
	G_{0,0} & G_{0,1} & \dots & G_{0,K}\\
	G_{1,0} & G_{1,1} & \dots & G_{1,K}\\
	\vdots & \vdots & \ddots & \\
	G_{K,0} & G_{K,1} & \dots & G_{K,K}\\ 
	\end{bmatrix}
\end{align}
is no longer circulant, but fully populated and dense.

\section{\label{sec:model} Plasma Simulation Models}

Harmonic balance can generally be applied to any plasma simulation that provides a voltage--current relation and operates in time-domain with steady-state. In this work, we use two different plasma models: a) A nonlinear global equivalent circuit model that can be coupled to a linear external circuit. b) A self-consistent 1-D PIC simulation to be coupled to a linear external circuit. Both models are used to simulate a capacitively coupled argon discharge at low pressure ($p < 10$~Pa).

In this work, only a single excitation frequency is considered. If the number of frequencies of interest need to be higher and these frequencies are harmonics of each other, the proposed method can be applied without any alterations. If this is not the case, the algorithm needs to be adjusted. A detailed discussion of this can be found in \cite{maas_nonlinear_2003}.

\subsection*{Global equivalent circuit model}

\begin{figure}[t!]
	\centering
	\resizebox{12cm}{!}{		
		\includegraphics[width=12cm]{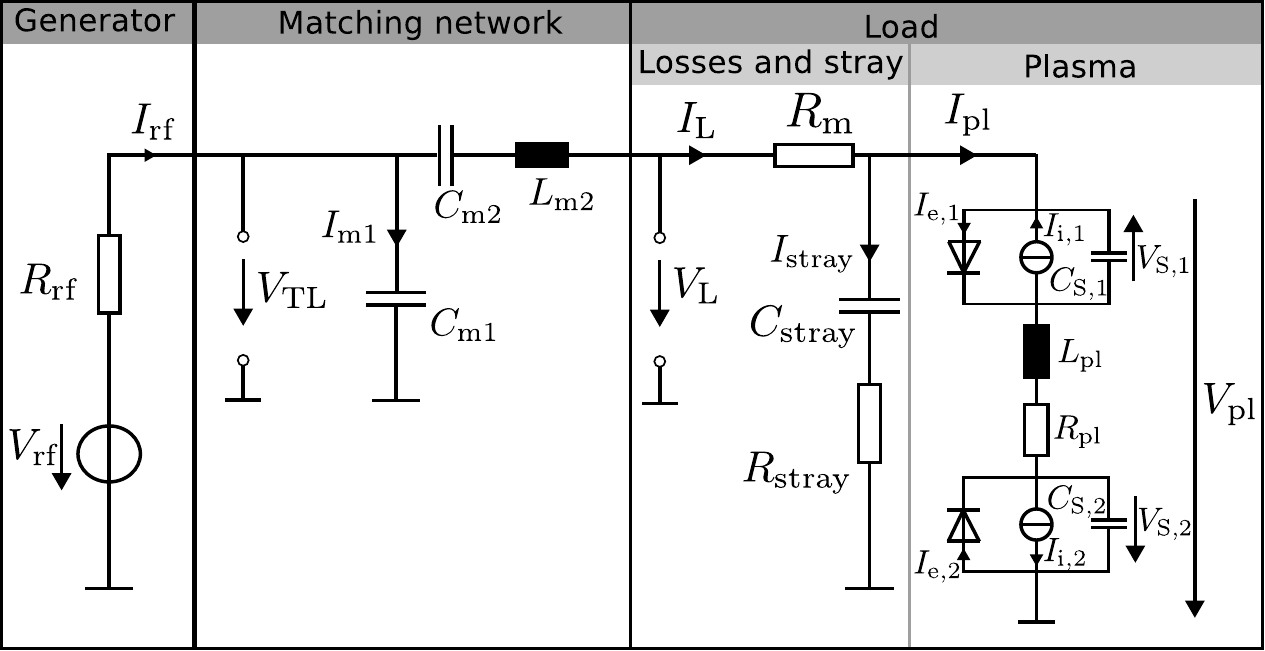}
	}
	\caption{Generator, matching network and stray elements attached to an equivalent circuit of the plasma.}
	\label{fig:setup}
\end{figure}

The global plasma model utilized in this work is based on considerations introduced and discussed in~\cite{mussenbrock_enhancement_2008, mussenbrock_nonlinear_2006, lieberman_effects_2008, ziegler_temporal_2009}, while its interaction with an external electric circuit has been studied using ngSPICE~\cite{vogt_ngspice_2017} in previous works \cite{schmidt_consistent_2018, schmidt_multi_2018}. In the following, the model is only briefly reviewed. For a more detailed description the referenced literature is suggested.

\begin{table}
\begin{tabular}{l  c}
\toprule
Parameter & Value \\
\colrule
$T_\mathrm{Ar}$ & 300 K \\
$n$ & $1.25 \times 10^{15}~\mathrm{m}^{-3}$ \\
$p$ & 0.66 Pa \\
$k_\mathrm{B} T_e$ & 4.73 eV \\
$A_\mathrm{E}$ & 100 cm$^2$\\
$A_\mathrm{G}$ & 300 cm$^2$ \\
$l_\mathrm{B}$ & 5.7 cm \\
$V_\mathrm{rf}$ & 100 V \\
$\omega$ & $2 \pi \times 13.56$~MHz \\
$R_\mathrm{rf}$ & $50~\Omega$ \\
$R_\mathrm{m}$ & $0.5~\Omega$ \\
$R_\mathrm{stray}$ & $0.5~\Omega$ \\
$C_\mathrm{stray}$ & $200$ pF \\
$C_\mathrm{m1}$ & $1550$ pF \\
$C_\mathrm{m2}$ & $175$ pF \\
$L_\mathrm{m2}$ & $1500$ nH \\

\botrule
\end{tabular}
\caption{Input parameters for the the global plasma model simulation.}
\label{table:parameter}
\end{table}

The model of the plasma is divided into two sheaths and a bulk. On the one hand, following a generalized Ohm's law, the bulk is modeled as an inductance $L_\mathrm{pl}=l_\mathrm{B} m_\mathrm{e}/e^2 n A_\mathrm{E}$ and a resistance $R_\mathrm{pl} = \nu_\mathrm{eff} L_\mathrm{pl}$, with the bulk length $l_\mathrm{B}$, the electron mass $m_\mathrm{e}$, the plasma density $n$, the electrode area  $A_\mathrm{E}$, and the effective collision frequency $\nu_\mathrm{eff}$. The sheaths, on the other hand, consist of a nonlinear capacitance, a constant current source to account for the steady ion flux and a diode to model the electron dynamics in the sheath. The current source has a value of $I_\mathrm{i,1} = A_\mathrm{E} e n u_\mathrm{B}$ for the driven electrode and $I_\mathrm{i,2} = A_\mathrm{G} e n u_\mathrm{B}$ for the grounded electrode with the grounded area $A_\mathrm{G}$, the Bohm velocity $u_\mathrm{B} = \sqrt{k_\mathrm{B} T_e/m_\mathrm{i}}$, the electron temperature $T_e$, and the ion mass $m_\mathrm{i}$. The electron current depends on the sheath voltage and amounts to $I_\mathrm{e,1} = A_\mathrm{E} e n \bar{v}_\mathrm{e} \mathrm{exp}\left( -e V_\mathrm{S,1}/k_\mathrm{B} T_\mathrm{e} \right)$ for the driven electrode and $I_\mathrm{e,2} = A_\mathrm{G} e n \bar{v}_\mathrm{e} \mathrm{exp}\left( -e V_\mathrm{S,2}/k_\mathrm{B} T_\mathrm{e} \right)$ for the grounded electrode, with mean electron speed $\bar{v}_\mathrm{e} = \sqrt{8 k_\mathrm{B} T_e/\pi m_\mathrm{i}}$. Lastly, the nonlinear capacitances have a value of $C_\mathrm{S,1}= \left( e n \epsilon_0 A_\mathrm{E}^2/2 V_\mathrm{S,1} \right)^{\frac{1}{2}}$ and $C_\mathrm{S,2}= \left( e n \epsilon_0 A_\mathrm{G}^2/2 V_\mathrm{S,2} \right)^{\frac{1}{2}}$, respectively, resulting from a Matrix sheath model \cite{lieberman_principles_2005}. Again, a detailed discussion about this model coupled to an external electrical circuit can be found elsewhere \cite{schmidt_consistent_2018}. The resulting equivalent circuit is depicted in Figure~\ref{fig:setup} on the far right side attached to a generator, a matching network and stray elements. An obvious method to simulate such a model with an external network attached is to make use of a circuit simulation tool such as ngSPICE \cite{vogt_ngspice_2017}. This model therefore serves as a proof of concept for the harmonic balance analysis, since it provides a reference method to simulate the setup. The values chosen for the plasma parameters and the network elements are the same as published in \cite{schmidt_consistent_2018} and are listed in Table~\ref{table:parameter}.

\subsection*{Particle in Cell}

For many investigations, a global plasma model is not satisfactory since all spatial information on the plasma is integrated and not resolved. For a detailed study of, e.g., the electron dynamics inside the sheath or the ion energy distributions at the walls, a spatially resolved simulation such as PIC is useful. PIC incorporates the plasma dynamics self-consistently through a kinetic description of electrons and ions coupled to the electromagnetic fields. As a result it offers more insights than a global model at the cost of being computationally expensive. In this work the 1-dimensional PIC code \textit{yapic} is used \cite{turner_simulation_2013, trieschmann_particle--cell/test-particle_2017}.

Two different setups are investigated: A geometrical symmetric setup with an electrode area of $A_\mathrm{E} = 315~\mathrm{cm}^{2}$ and a resistance and capacitance in series (RC-unit) attached to it. The values of the external circuit elements are $V_\mathrm{S} = 100$~V, $R=10~\Omega$, and $C = 300$~pF. This simple circuit can be included by making use of harmonic balance, but alternatively also by solving the network's auxiliary differential equations simultaneously with the discharge as proposed by Verboncoeur et al. \cite{verboncoeur_simultaneous_1993}. Similar to the simulation of the global plasma model, this approach serves as a reference for a proof of concept, since two methods for solving the same case are available. The second setup adds a more realistic external circuit to the PIC simulation, namely, the one depicted in Figure~\ref{fig:setup} with the plasma being modeled by PIC instead of the equivalent circuit. A geometrically asymmetric discharge in spherical coordinates is simulated with a driven electrode area $A_\mathrm{E} = 50~\mathrm{cm}^{2}$ and a grounded electrode area $A_\mathrm{G}=1250~\mathrm{cm}^{2}$. For this network no reference simulation exists, making this a demonstration of the flexibility provided by harmonic balance.
Both cases use an argon discharge with a pressure $p=1$~Pa and a temperature of 650~K.

The simulations with respectively varied voltages are always initiated with the same steady-state solution, i.e. the same number of particles and their distribution in phase-space. The assumption is that the voltages and thereby the state of the system does not change significantly. In this case, convergence can be reached faster.

\section{\label{sec:results} Results and Discussion}

\subsection*{Global equivalent circuit model}

\begin{figure}[t!]
	\centering
	\resizebox{12cm}{!}{		
		\includegraphics[width=12cm]{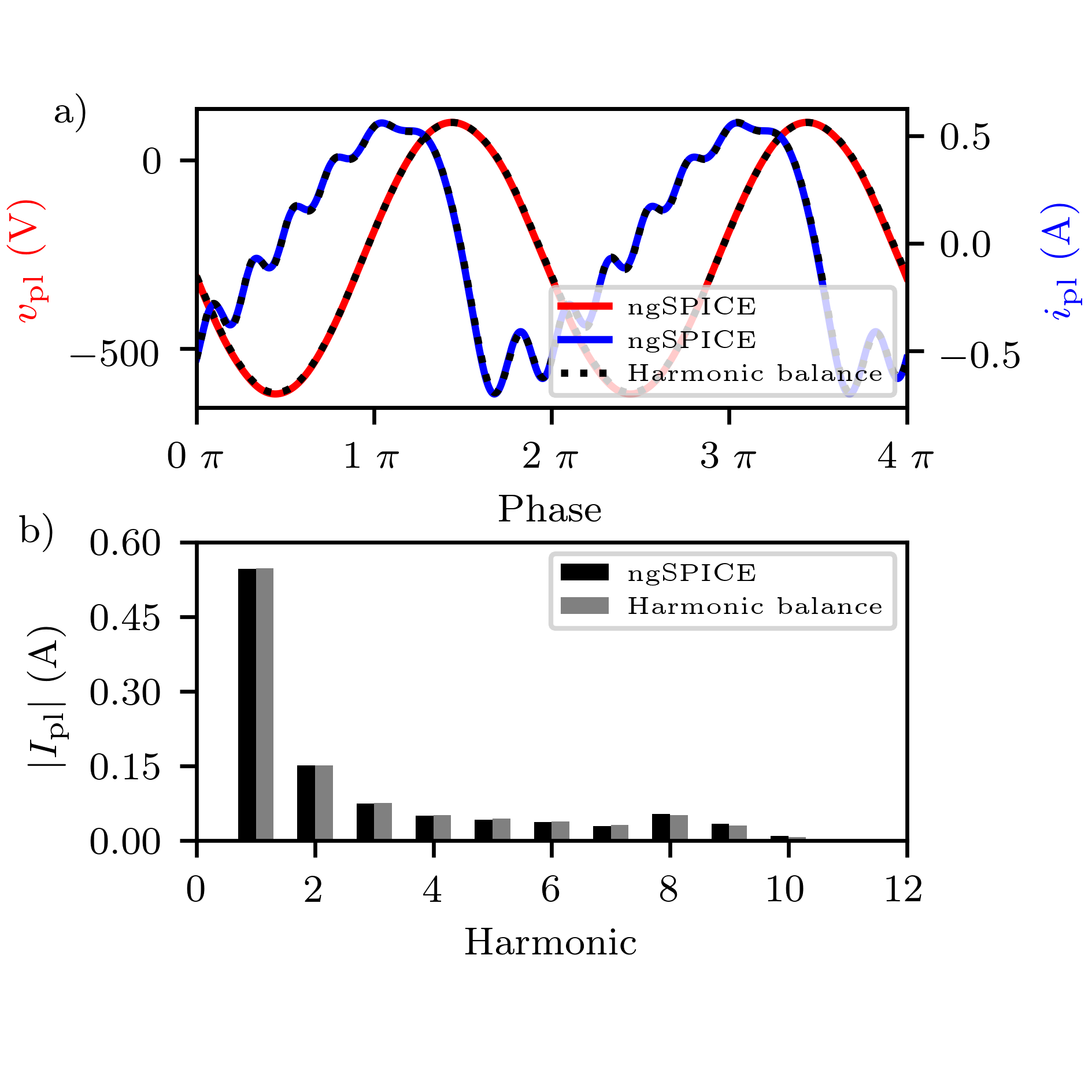}
		
	}
	\caption{Voltage and current using the global plasma model. a) Transient solution. The straight lines indicate the results obtained using ngSPICE, while the dotted lines represent the results obtained using the harmonic balance algorithm. b) The absolute values of the different current harmonics. In black are the results obtained by ngSPICE, in grey are those which are obtained from the harmonic balance algorithm.}
	\label{fig:result_global}
\end{figure}

The network depicted in Figure~\ref{fig:setup} is simulated with both ngSPICE and the harmonic balance algorithm. Within the latter, the whole global plasma model is treated as the nonlinear part, while the rest of the circuit (excluding the voltage source) is incorporated in the transadmittance matrix. Thereby, the solution of the plasma model can conceptually be obtained using any arbitrary method, such as ngSPICE or Mathematica. Performing the steps described in Section~\ref{sec:method}, a converged solution gives the steady-state voltages and currents in the system. The transient values of $v_\mathrm{pl}(t)$ and $i_\mathrm{pl}(t)$ (corresponding to Fourier components $V_\mathrm{pl}$ and $I_\mathrm{pl}$) are shown in Figure~\ref{fig:result_global}. In a) the result of the harmonic balance method is plotted together with a reference simulation using ngSPICE, while b) shows the corresponding Fourier components of the current $I_\mathrm{pl}$. It is obvious that both results are practically identical, which is the desired outcome. The differences that still arise can be explained with numerical inaccuracies and a finite number of $K=15$ considered harmonics in the Fourier series. While the voltage is almost completely sinusoidal at the fundamental excitation frequency, the current consists of a number of harmonics due to the strong interaction of the plasma bulk and the nonlinear sheaths at low pressure and given the asymmetry of the setup. More details on the discharge physics and the matching procedure are discussed in \cite{schmidt_consistent_2018}.

\subsection*{Particle in Cell}

\begin{figure}[t!]
	\centering
	\resizebox{12cm}{!}{		
		\includegraphics[width=12cm]{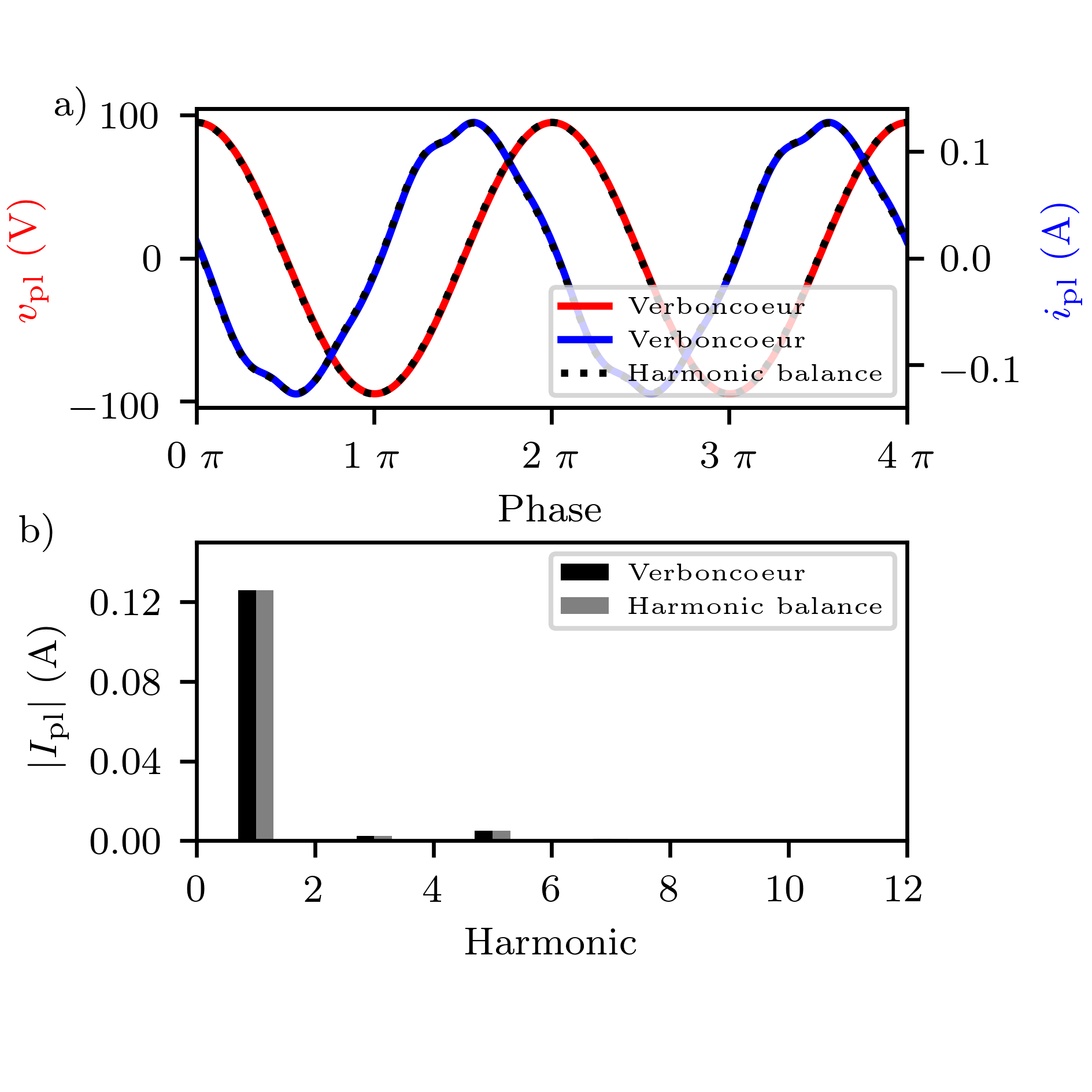}
		
	}
	\caption{Plasma-voltage and current using PIC and an external RC-element. a) Transient solution. The straight lines indicate the results using the method proposed by Verboncoeur et al.\ \cite{verboncoeur_simultaneous_1993}, while the dotted plots show the results obtained using the harmonic balance algorithm, b) The absolute values of the different harmonics of the current. In black are the results obtained using the method proposed by Verboncoeur et al.\ \cite{verboncoeur_simultaneous_1993}, in grey are those obtained from the harmonic balance algorithm.}
	\label{fig:result_PIC_RC}
\end{figure}

First, the symmetric setup with an RC-unit attached and solved using PIC simulations is considered. The current flowing through the discharge is expected to contain only a small amount of harmonics aside from the excitation frequency. Figure~\ref{fig:result_PIC_RC} shows the simulation result for the plasma current and voltage in which the external circuit is simulated using the harmonic balance approach and the method proposed by Verboncoeur et al.\ \cite{verboncoeur_simultaneous_1993}. Both cases lead to the same result of a sinusoidal voltage of 95~V amplitude and a current with 120~mA amplitude and small amounts of higher harmonics. It is worth noting that even in a  symmetric arrangement of the discharge higher harmonics can be observed, since the nonlinear characteristics of the two opposing sheaths do not completely cancel. In terms of a Taylor expansion of the voltage charge characteristics of the sheaths, only the even series elements cancel. The odd elements remain. This is the reason why one can observe only odd harmonics in the current in the case of a perfectly symmetric discharge. 

\begin{figure}[t!]
	\centering
	\resizebox{12cm}{!}{		
		\includegraphics[width=12cm]{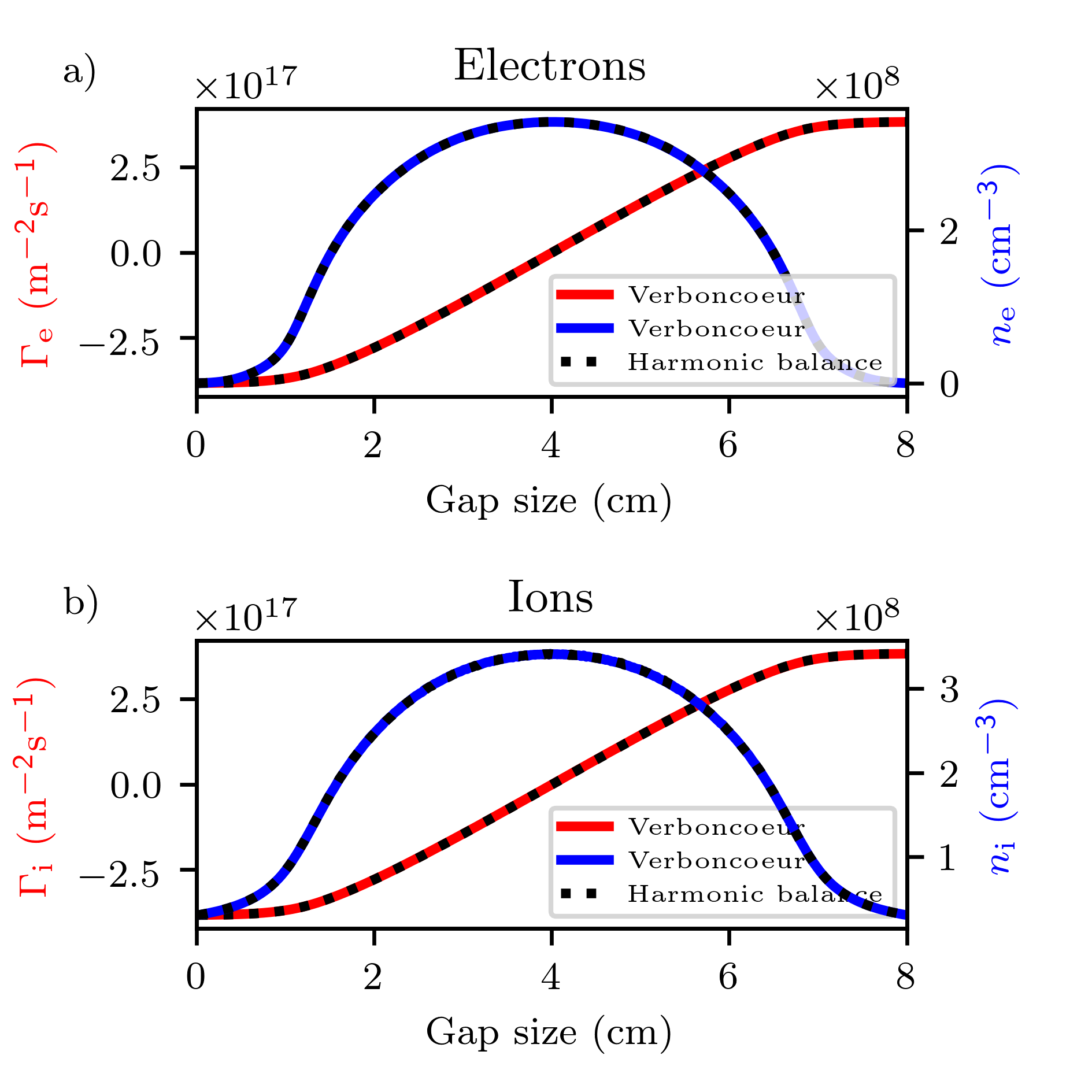}
		
	}
	\caption{Densities of electrons and ions and flux of the particles resulting from a PIC simulation with an external RC-element. Straight lines indicate the solutions obtained from the method proposed by Verboncoeur et al.\ \cite{verboncoeur_simultaneous_1993}, dotted lines show the results from the harmonic balance algorithm. a) Density and flux of electrons. b) Density and flux of ions. }
	\label{fig:result_PIC_RC_densities}
\end{figure}

Using these results, the plasma impedance can be calculated at the excitation frequency to $Z_\mathrm{pl} =(53-j 760)~\Omega$. The external circuit has an impedance of $Z_\mathrm{ext} = (10 - j 39)~\Omega$. Taking these impedances as a voltage divider and with the source having an amplitude of 100 V, the voltage drop at $Z_\mathrm{pl}$ is calculated to 95 V, which is consistent with the obtained simulation results.

The average ion and electron densities and fluxes depicted in Figure~\ref{fig:result_PIC_RC_densities} a) and b) are also identical within the level of statistical accuracy for both simulation methods. This proves that not only the global voltage and current are the same, but also the intrinsic plasma state, including the spatio-temporal dynamics.

\begin{figure}[t!]
	\centering
	\resizebox{12cm}{!}{		
		\includegraphics[width=12cm]{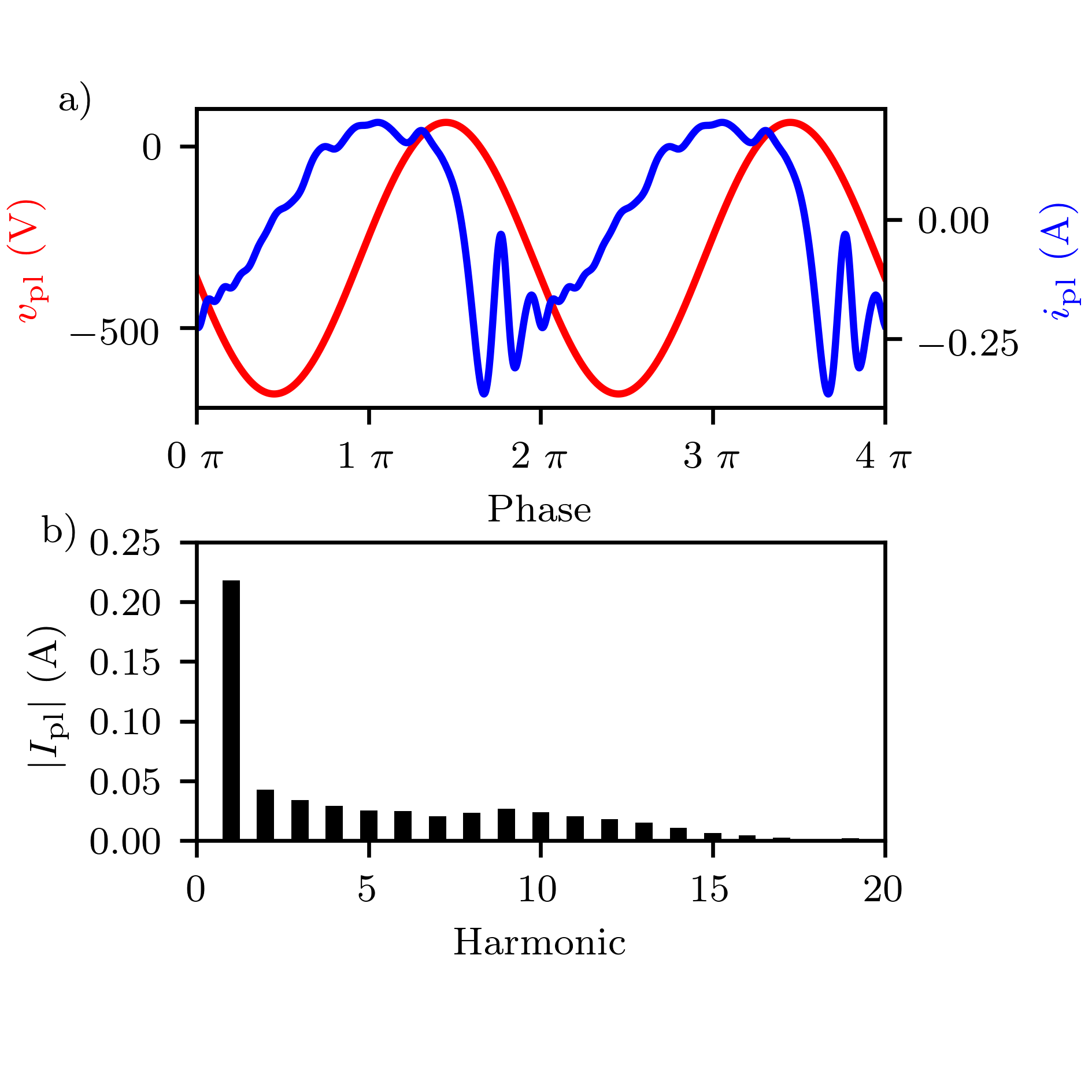}
		
	}
	\caption{Current and voltage resulting from a PIC simulation with an attached generator, matching network and stray elements. a) Transient solution b) Absolut values of different harmonics in the current}
	\label{fig:result_PIC_match}
\end{figure}

The second case with an attached generator, the matching network and reactor losses was simulated accordingly. The voltage and current evolution is depicted in Figure~\ref{fig:result_PIC_match}. As expected for an asymmetric setup, the current is very nonlinear, containing multiple harmonics. At the same time, the voltage remains almost completely sinusoidal. The results shown are for a matched case, which means that the impedance seen by the generator at the fundamental frequency is $Z_\mathrm{TL}=V_\mathrm{TL}/I_\mathrm{TL} \approx 50~\Omega$. This is achieved by iteratively varying the capacitances $C_\mathrm{m1}$ and $C_\mathrm{m2}$ until matching is obtained \cite{schmidt_consistent_2018}. The resulting values are $C_\mathrm{m1}=1536$~pF and $C_\mathrm{m2}=185$~pF -- due to the matching to a different load -- while all other network elements have the same value as listed in Table~\ref{table:parameter}.

\section{\label{sec:conclusion} Conclusion}

A solution scheme is developed, which allows for the coupling of arbitrary lumped element circuits to radio-frequency plasma simulations. The approach is based on the harmonic balance method and incorporates the external circuit via a simple netlist, avoiding the deployment and the solution of Kirchhoff's differential equations by hand. 

The validity of the proposed simulation approach is demonstrated using two different reference simulations. First, a global plasma model is established and attached to an external matching circuit. The results obtained by harmonic balance are identical to a simulation based on the electrical network analysis tool ngSPICE. Second, a geometrically symmetric 1-dimensional PIC simulation is coupled to a resistance and a capacitance in series (i) via conservation of charge at the driven electrode and a coupling of the circuit equations to the PIC simulation by hand and (ii) using harmonic balance. Also in this case, the results are practically indistinguishable.

Lastly, a more complicated electrical network consisting of a generator, a matching network and stray elements is connected to PIC using harmonic balance to demonstrate the versatility of the method also for cases where the circuit is not straightforwardly incorporated via auxiliary differential equations.

The presented method offers a fast and comfortable solution for the integration of complex external networks into plasma simulations.
Depending on the nonlinearity of the considered discharge or with voltage source contributions that are not harmonics of the fundamental frequency, a high number of harmonics in the current may result and, consequently, an equivalent number of simulations have to be considered.
This extensive necessity may ultimately lead to critical computational costs. In the future, an analysis of the latter and possibly an optimization might therefore be inevitable.

\section*{Acknowledgement}

This work is supported by the German Research Foundation (DFG) in the frame of Transregional Collaborative Research Centre TRR\,87, Collaborative Research Centre SFB\,1316, and DFG Research Grant MU2332/6-1. 

\section*{ORCID IDs}
\noindent\href{https://orcid.org/0000-0001-6623-0464}{F. Schmidt: https://orcid.org/0000-0001-6623-0464}

\noindent\href{http://orcid.org/0000-0001-6445-4990}{T. Mussenbrock: http://orcid.org/0000-0001-6445-4990}

\noindent\href{http://orcid.org/0000-0001-9136-8019}{J. Trieschmann: http://orcid.org/0000-0001-9136-8019}

\noindent\href{http://orcid.org/0000-0001-5041-2941}{T. Gergs: http://orcid.org/0000-0001-5041-2941}

\bibliographystyle{aip}

\bibliography{Harmonic_Balance.bib}

\end{document}